\documentclass[10pt,journal,letterpaper]{IEEEtran}
\IEEEoverridecommandlockouts
\usepackage[utf8]{inputenc}
\usepackage[T1]{fontenc}
\usepackage{orcidlink}
\usepackage{setspace}
\usepackage{booktabs}
\usepackage{tabularx}
\IEEEoverridecommandlockouts
\usepackage{fancyhdr}
\usepackage[numbers, sort&compress]{natbib}
\usepackage{url}
\usepackage{graphicx}

\usepackage{float}
\usepackage{enumerate}
\usepackage{hyperref}
\usepackage{amsmath,amssymb,amsfonts}
\usepackage{algorithmic}
\usepackage[ruled,vlined]{algorithm2e}
\usepackage{graphicx}
\usepackage{array}     
\usepackage{amsmath}   
\usepackage{textcomp}
\usepackage{xcolor}
\usepackage{physics}
\usepackage[caption=false]{subfig}
\usepackage[font=footnotesize, skip=0pt,compatibility=false]{caption}
\usepackage{subcaption}
\usepackage{comment}
\usepackage{tikz}
\usetikzlibrary{positioning, shapes, arrows.meta}
\usepackage{booktabs}
\usepackage{multirow}
\def\BibTeX{{\rm B\kern-.05em{\sc i\kern-.025em b}\kern-.08em
    T\kern-.1667em\lower.7ex\hbox{E}\kern-.125emX}}
\begin{document} 
\bstctlcite{MyBSTcontrol}
\title{Synchronized Quantum Devices in Lossy Channels}
\author{Ravi Singh Adhikari$^{*}$, Aman Gupta$^{*}$, Xiaoyu~Ai$^{*}$, and~Robert~Malaney$^*$ 
\thanks{$^*$Ravi Singh Adhikari, Aman Gupta,   Xiaoyu Ai, and Robert Malaney are with the School of Electrical Engineering and Telecommunications, University of New South Wales, Sydney, Australia.}
}

\maketitle
\thispagestyle{empty}
\pagestyle{empty}
\begin{abstract}

Due to their tight temporal coupling, entangled photon pairs produced by spontaneous parametric down-conversion offer a new route to improved network synchronization among quantum-enabled devices. In the emerging space-based quantum internet, such synchronization will be pivotal to operational success. Here, we propose and experimentally demonstrate a new polarization-assisted  protocol that allows quantum-enabled devices to remain synchronized under the high loss conditions  anticipated for satellite-to-ground quantum channels. 
Our experimental results show that polarization assistance extends reliable synchronization acquisition to higher channel losses and an increase of up to 40$\%$ in satellite-to-ground distance.  
 Supporting Monte Carlo simulations show that this performance improvement extends over a wide range of photon-starved conditions. Collectively, these results not only provide a proof of concept for polarization-assisted network synchronization but also demonstrate its pragmatic importance in significantly extending the range under which quantum-enabled devices forming the backbone of the quantum internet can remain highly synchronized.


\end{abstract}

\section{Introduction}\label{sec:introduction}
Satellite-based quantum communication is emerging as a key component of future secure communication networks, for which precise synchronization between remote clocks is an important enabling capability. More broadly,  clock synchronization, the estimation and compensation of both clock offset and clock-drift rate, is beneficial for distributed quantum computing~\cite{PhysRevA.59.4249}, quantum networking~\cite{xiang2025towards}, telecommunications~\cite{pini2021satellite}, satellite-based quantum communications~\cite{yin2017satellite1,Dai2020}, and precise navigation and positioning~\cite{surof2026precise}.  
During the past decade, quantum clock synchronization (QCS) protocols~\cite{Lee2019,PhysRevA.100.023849,Quan:22} have been investigated, along with quantum-secure time transfer protocols that integrate high-precision synchronization with quantum-security mechanisms~\cite{Dai2020,adhikari2025new}. 
In principle, entangled photon-based QCS protocols can achieve sub-picosecond-scale precision in clock synchronization by exploiting the tight birth-time correlation of entangled photon pairs~\cite{quan2018experimental, PhysRevApplied.19.054082}. 

In ideal low-background environments, true pair coincidences\footnote{A  ``pair coincidence'' refers to detections in separate receivers within a given timing window. The qualifier ``true'' denotes detections  from the same entangled photon pair, while ``accidental'' denotes all other coincidence events.} produce a narrow temporal-correlation peak in the distribution of photon-arrival time differences~\cite{Ho_2009}. The peak position of this distribution provides an estimate of the clock offset, while variations in successive peak positions can be used to estimate the clock-drift rate. QCS protocols based on this technique have been demonstrated in laboratory settings and fiber links~\cite{Lee2019,  PhysRevA.100.023849,  Quan:22, PhysRevApplied.22.024012, PhysRevApplied.19.054082, Pelet2025}. However, satellite channels face significant practical challenges. A typical satellite-to-ground channel is affected by geometric loss, receiver loss, atmospheric turbulence, pointing errors, detector dark counts, and solar background photons, collectively resulting in a low detected signal-to-noise ratio~\cite{yin2017satellite1, lu2022micius, lafler2023quantumtimetransferpractical}.

For typical aperture sizes (at the satellite and ground station) anticipated for deployed satellite channels, 
combined losses in the channel can be in the $50$dB--$60$dB range~\cite{ sidhu2021advances,lu2022micius}, and at these losses the true pair coincidence rate becomes comparable to or smaller than the accidental pair coincidence rate. We refer to a signal of this form as being in the `photon-starved' regime. In this regime, the temporal-correlation peak becomes increasingly difficult to distinguish from the accidental pair coincidence background~\cite{Ho_2009}, ultimately preventing reliable synchronization acquisition.
Alleviating this undesirable outcome is the focus of this work.
 
Previous QCS protocols that utilized pairs of entangled polarized photons relied on birth-time correlations to achieve precise clock synchronization~\cite{ marcikic2006free,10.1063/1.5121489, quan2018experimental}. Polarization correlations have also been used for verification of entanglement and synchronization security~\cite{alqedra2025entanglementverifiedtimedistributionmetropolitan,adhikari2025new, adhikari2026quantumsecuretimetransfer}. In polarization-encoded quantum key distribution protocols, polarization information has been used for secret-key generation, while polarization filtering has also been investigated for background suppression~\cite{castillo2023experimental,simmons2024dawn}.
However, none of these protocols has directly linked the polarization correlation to the synchronization protocol itself. Here, we demonstrate how exploiting the polarization correlation to suppress accidental pair coincidences during clock synchronization improves synchronization acquisition under photon-starved conditions. Specifically, we find that embedding this new technique into a synchronization protocol leads to reliable single-photon network synchronization over satellite-ground distances of up to 40$\%$ more than otherwise possible, a significant improvement that has important ramifications for emerging quantum networks.

\begin{figure*}[htp] 
    \centering
    \includegraphics[width=1\linewidth]{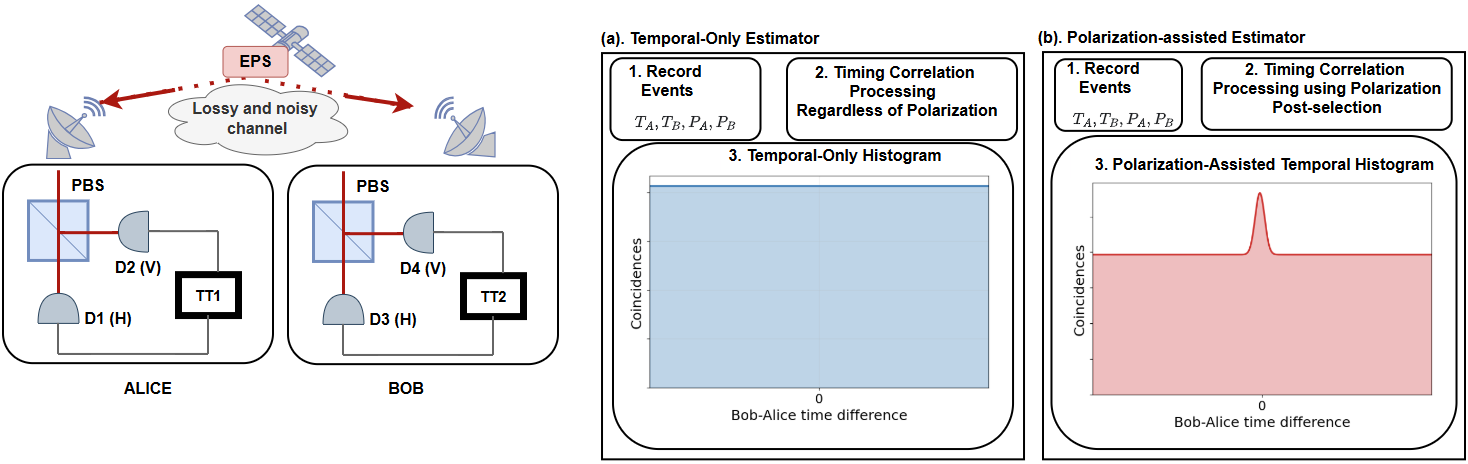}
    \vspace{4pt} 
    \caption{Temporal-only and polarization-assisted estimators using identical polarization-resolved receiver hardware. An entangled-photon source (EPS) onboard the satellite distributes correlated photon pairs to two ground stations through independent lossy and noisy free-space channels. At each receiver, a polarizing beam splitter (PBS) separates the incoming photons into horizontal (H) and vertical (V) components, which are detected by polarization-resolved single-photon detectors (D1--D4) and recorded by local time taggers (TT1 and TT2). (a) In the temporal-only estimator, the polarization arrays $P_A$ and $P_B$ are discarded, and the temporal-correlation histogram is constructed from the recorded time-tag arrays $T_A$ and $T_B$, irrespective of polarization. (b) In the polarization-assisted estimator, the same recorded data are processed using polarization post-selection, retaining pair coincidences satisfying the polarization-matching condition before the construction of the temporal-correlation histogram. The two estimators therefore use the same source, channel conditions, detectors, and time-tagging hardware, so that any improvement in the correlation peak arises solely from exploiting the recorded polarization information.}
\label{fig. method}
   \end{figure*}
   
The remainder of this paper, and its main contributions, can be summarized as follows.
\begin{enumerate}
    \item We formulate a statistical model for pair coincidences and predict the coincidence-to-accidental ratio (CAR) enhancement provided by our polarization post-selection technique. A comparison with the enhancement derived from Fisher-information analysis is provided.
    \item We then \textit{experimentally} demonstrate polarization-assisted clock synchronization using a polarization-entangled photon source and verify that the predicted enhancement in synchronization-acquisition performance is achieved.
    
    \item Finally, we perform supporting Monte Carlo simulations over a broad range of channel-loss and background-count conditions, showing that our experimentally observed advantage will persist across a wide range of operational photon-starved conditions.
\end{enumerate}

\section{Polarization-Assisted Clock Synchronization}
\label{sec:method}
\subsection{System Model}
As shown in Fig. \ref{fig. method}, a satellite equipped with an onboard entangled-photon source distributes one photon from each photon pair to a ground station, Alice, and the other photon to a ground station, Bob; each station is equipped with a photon-detection setup and a local clock.
Let the clock offset between Alice and Bob be~$\tau=\tau_B^o-\tau_A^o$, where~$\tau_A^o$ and~$\tau_B^o$ denote the offsets of Alice and Bob's local clocks, respectively, relative to a common reference O. The detection time of a photon from a true photon pair at the ground station, $g$, is given by~$t_g= t^o+{L_g}/{c} + \tau^o_g +\epsilon_g,~\; {g} \in \{A, B\}$, where $g$ distinguishes Alice and Bob using labels $A$ and $B$, respectively, $t^o$ is the photon-pair emission time relative to O, $L_g$ is the distance from the satellite to $g$, and $\epsilon_g$ represents the timing error associated with detection at $g$. Assuming that satellite-to-ground propagation distances are accurately known, we compensate for the propagation-delay difference and define~$\Delta t=t_B-t_A-\frac{L_B-L_A}{c}=\tau+\epsilon$, where 
~$\epsilon=\epsilon_B-\epsilon_A$. Assuming independent Gaussian timing errors at Alice's station and Bob's station, $\epsilon_A\sim\mathcal{N}(0,\sigma_A^2)$ and $\epsilon_B\sim\mathcal{N}(0,\sigma_B^2)$, respectively, where $\sigma_A$ and $\sigma_B$ are the corresponding standard deviations (SDs). Since $\epsilon=\epsilon_B-\epsilon_A$, then $\epsilon\sim\mathcal{N}(0,\sigma_{\mathrm{sys}}^2)$, where $\sigma_{\mathrm{sys}}^2=\sigma_A^2+\sigma_B^2$. Throughout this work, the notation \(\sigma_h\) represents the SD of a variable $h$.

\subsection{Temporal and Polarization Statistics}
We model each pair coincidence as either a true pair coincidence or an accidental pair coincidence. We use the subscripts \(t\) and \(a\) to label quantities associated with true and accidental pair coincidences, respectively. 

\subsubsection{Temporal statistics}
We consider a Bell state~$|\Phi^+\rangle=\dfrac{1}{\sqrt{2}}(|\text{H}\rangle_{\text{S}} |\text{H}\rangle_{\text{I}} + |\text{V}\rangle_{\text{S}} |\text{V}\rangle_{\text{I}})$, with polarization visibility, $\mathcal{V}\in[0,1]$~\cite{anwar2021entangled}. Here, $\text{H}$ and $\text{V}$ represent the horizontal and vertical polarization of the photon; the subscripts $\text{S}$ and $\text{I}$ denote the signal and idler, respectively. 

The propagation-delay-compensated timing probability density function (PDF) of a true pair coincidence is given by
$
\mathcal{N}(\Delta t| \tau, \sigma_{\mathrm{sys}}^2 )=
\frac{1}
{\sqrt{2\pi}\sigma_{\mathrm{sys}}}
\exp
\left[
-\frac{(\Delta t-\tau)^2}
{2\sigma_{\mathrm{sys}}^2}
\right]$.
An accidental pair coincidence arises from independent detection events, including detector dark counts, background counts, and detections originating
from different photon-pair arrival events.
We model the timing PDF for an accidental pair coincidence as uniform over a timing window, $\Delta \mathcal{T}$, such that
$\mathcal{U}(0, \Delta \mathcal{T})=1/\Delta \mathcal{T},~\Delta t\in [0,\Delta \mathcal{T}]$. 
Note that $\Delta \mathcal{T}$ is assumed to be sufficiently wide relative to \(\sigma_{\rm sys}\) so that the truncation of the Gaussian true-coincidence peak is negligible.

\subsubsection{Polarization Statistics}
Let $p$ denote a discrete joint polarization outcome, \(p\in\Omega_p\) associated with a pair coincidence, where \(\Omega_p=\{\mathrm{HH},\mathrm{VV},\mathrm{HV},\mathrm{VH}\}\).
We define the set of polarization-matched outcomes as~$\Omega_{\rm PM}=\{\mathrm{HH},\mathrm{VV}\}$,
where the subscript \({\rm PM}\) denotes polarization matching.
Assuming a symmetric polarization error in the H/V measurement basis,
the probability of  a joint polarization outcome $p$ for a true pair coincidence is
\[
\text{Pr}_t(p)
=
\begin{cases}
(1+\mathcal{V})/4, & p\in \Omega_{\rm PM} ,\\
(1-\mathcal{V})/4, & p\notin \Omega_{\rm PM}  .
\end{cases}\]
The probability that a true pair coincidence satisfies the PM condition is therefore~$q_t=\sum\limits_{p\in\Omega_{\rm PM}}\text{Pr}_t(p)=\frac{1+\mathcal{V}}{2}$. For unit visibility, $\mathcal{V}=1$, and thus $q_t=1$. Imperfect polarization visibility causes a fraction of true pair coincidences to appear in mismatched outcomes $(\text{H}\text{V},\text{V}\text{H})$. 

For accidental pair coincidences, we assume that Alice's and Bob's polarization outcomes are uncorrelated. The probability of a joint polarization outcome $p$ for an accidental pair coincidence is $\text{Pr}_a(p)=\frac{1}{4}$. The probability that an accidental pair coincidence satisfies the PM condition is $q_a=\frac{1}{2}$. 
We further assume that the temporal statistics are independent of the polarization statistics. Thus, polarization post-selection changes the relative numbers of retained true pair coincidences and accidental pair coincidences, but not their timing PDFs, $\mathcal{N}(\Delta t| \tau, \sigma_{\mathrm{sys}}^2 )$ and $\mathcal{U}(0, \Delta \mathcal{T})$.

\subsection{Clock-Offset Estimators}
\subsubsection{Temporal-only estimator}
Let \(\bar N_t \) and \(\bar N_a\) denote the expected numbers of true and accidental pair coincidences, respectively, within $\Delta \mathcal{T}$. The total expected number of pair coincidences used by the temporal-only estimator is given by~$\bar N_T=\bar N_t+\bar N_a$, where the subscript \(T\) denotes the temporal-only estimator. Let the true pair coincidence fraction be given by~$\rho=\frac{\bar N_t}{\bar N_t+\bar N_a}$, where $0\le\rho\le1$. 
Hence, the timing PDF used by the temporal-only estimator is the Gaussian--uniform mixture given by 
\begin{equation}\label{timing-only}
    f_T(\Delta t)
=
\rho \mathcal{N}(\Delta t| \tau, \sigma_{\mathrm{sys}}^2 )
+
(1-\rho)\mathcal{U}(0, \Delta \mathcal{T}).
\end{equation}

The temporal-only estimator uses the recorded photon-arrival times while discarding the associated polarization outcomes. For \(\bar N_T \) pair coincidence timing differences, $\{\Delta t_n\}_{n=1}^{\bar N_T}$,
the log-likelihood is~$\ell_T(\tau)
=
\sum_{n=1}^{\bar N_T}
\ln f_T(\Delta t_n ).$ The corresponding maximum-likelihood estimate of the clock offset is  $\hat{\tau}_T=\operatorname*{arg\,max}_{\tau}
\ell_T(\tau).$
\subsubsection{Polarization-assisted estimator}
For the target state $|\Phi^+\rangle$, we define the polarization-matching function $\mathrm{M}(p)$ as:
\[\mathrm{M}(p)
=
\begin{cases}
1,
&
p\in \Omega_{\rm PM},
\\[4pt]
0,
&
p\notin \Omega_{\rm PM}.
\end{cases}\]
The polarization-assisted estimator first applies \(\mathrm{M}(p)\) and retains only pair coincidences satisfying \(\mathrm{M}(p)=1\).
We refer to this selection as polarization post-selection.
After such selection, the expected numbers of retained pair coincidences are $\bar{N}_{t,P}=q_t\bar{N}_t,$ and $\bar{N}_{a,P}=q_a\bar{N}_a$, respectively, where the subscript \(P\) denotes the polarization-assisted estimator. 
Therefore, the expected total number of retained pair coincidences is $\bar{N}_P=q_t\bar{N}_t+q_a\bar{N}_a$. The true pair coincidence fraction therefore becomes~$\rho_P=\frac{q_t\bar{N}_t}{q_t\bar{N}_t+q_a\bar{N}_a}$. The corresponding polarization-assisted timing PDF, $f_P(\Delta t)$, is given by Eq.~\ref{timing-only} with $\rho$ replaced by~$\rho_P$;~$f_P(\Delta t)
=
\rho_P \mathcal{N}(\Delta t| \tau, \sigma_{\mathrm{sys}}^2 )
+
(1-\rho_P)\mathcal{U}(0, \Delta \mathcal{T})$. 
For \(\bar N_P \) pair coincidence timing differences, $\{\Delta t_n\}_{n=1}^{\bar N_P}$,
the polarization-assisted log-likelihood is
$\ell_P(\tau)
=
\sum_{n=1}^{\bar N_P}
\ln f_P(\Delta t_n),$
and the corresponding estimator is
$\hat{\tau}_{P}
=
\operatorname*{arg\,max}_{\tau}
\ell_P(\tau).$
The log-likelihood formulation provides the statistical model used for the analytical characterization, while the practical implementation estimates the clock offset from the peak of the correlation histogram~\cite{ PhysRevApplied.19.054082}.

For Alice's and Bob's time tags, $t_{A,i}$ and $t_{B,j}$, respectively, define the propagation-delay-compensated timing difference as~$d_{ij}=t_{B,j}-t_{A,i}-\frac{L_B-L_A}{c}$.
We define the practical polarization-assisted histogram estimator as,
$C_P(k)
=
\sum_{i=1}^{S_A}
\sum_{j=1}^{S_B}
M\!\left(p_{ij}\right)
\mathbf{1}
\!\left[
d_{ij}\in\mathcal{B}_k
\right],$ where $i$ and $j$ index Alice's and Bob's time tags, respectively; $S_A$ and $S_B$ are the
single count rates at Alice and Bob's stations, respectively,
 and \(p_{ij}\) is the corresponding joint polarization outcome associated with Alice's \(i\)-th and Bob's \(j\)-th detections.
The indicator $\mathbf{1}[d_{ij}\in\mathcal B_k]$ equals one when the timing difference $d_{ij}$ lies within bin $\mathcal B_k$ associated with bin index $k$, and zero otherwise. The polarization-assisted clock-offset estimate is obtained by first
selecting the timing bin with the maximum histogram count,
$k^\star
=
\operatorname*{arg\,max}_{k} C_P(k),$ and then assigning the center of the selected bin as the estimated clock offset, $\hat{\tau}_P
= \tau_{k^\star}.$ Here, $\tau_{k^\star}$ denotes the center of the timing bin
corresponding to $k^\star$. Likewise, the temporal-only histogram estimator is given by~$C_T(k)
=
\sum_{i=1}^{S_A}
\sum_{j=1}^{S_B}
\mathbf{1}
\!\left[
d_{ij}\in\mathcal{B}_k
\right].$ 
The two practical estimators therefore operate on the same polarization-resolved detection data; the polarization-assisted estimator differs only by the inclusion of the polarization-matching factor $\mathrm{M}(p_{ij})$. Given these two estimators, the clock-drift rate is derived from the slope of the successive clock-offset estimates over time.

\subsection{Analytical Performance Interpretation}
\subsubsection{Coincidence-to-Accidental Ratio Enhancement} 
Here, we define the CAR as the ratio of the expected number of true pair coincidences to that of accidental pair coincidences.
Before polarization post-selection, the CAR is $\text{CAR}_T=\frac{\bar N_t}{\bar N_a}$. After polarization post-selection, the CAR becomes~$\text{CAR}_P=\frac{q_t\bar N_t}{q_a\bar N_a}$. Therefore, the ratio of the polarization-assisted CAR to the temporal-only CAR is $\frac{\text{CAR}_P}{\text{CAR}_T}=\frac{q_t}{q_a}$. For the unpolarized-background case, $q_a=\frac{1}{2}$. Using $q_t=\frac{1+\mathcal{V}}{2}$, the CAR-enhancement factor becomes~$\frac{\text{CAR}_P}{\text{CAR}_T}=1+\mathcal{V}\geq1$. For unit visibility, $\mathcal{V}=1$, the CAR is enhanced by a factor of two, corresponding to~$10\log_{10}(2) \simeq 3\,\mathrm{dB}$.
\vspace{1em}

\subsubsection{Classical Fisher Information and Cram\'er-Rao Lower Bound}
We quantify the lower bound on clock-offset estimation uncertainty using the classical Fisher information and Cram\'er--Rao lower bound (CRLB)~\cite{steven1993fundamentals}. We treat all parameters other than~$\tau$ as known.

 \textit{(a) Temporal-only estimator}: 
    The Fisher information associated with $\bar N_T$ pair coincidences is given by
\begin{equation}\label{FIT}
    I_T = \bar N_T \int_{0}^{\Delta \mathcal{T}} f_T(\Delta t)
\left[
\frac{\partial}{\partial \tau}
\ln f_T(\Delta t)
\right]^2
\, d\Delta t.
\end{equation}
Since \(\mathcal{U}\) is independent of $\tau$, we have~$\frac{\partial f_T}{\partial\tau}
=
\rho\frac{\partial \mathcal{N}}{\partial\tau}
=
\rho
\frac{\Delta t-\tau}{\sigma_{\mathrm{sys}}^2}
\mathcal{N}$.
Substituting the expression of $\frac{\partial f_T}{\partial \tau}$ in Eq.~\ref{FIT} gives:
\begin{equation}\label{final FIT}
I_T
=
\bar N_T
\int_{0}^{\Delta \mathcal{T}}
\frac{
\rho^2
(\Delta t-\tau)^2
\mathcal{N}^2
}{
\sigma_{\mathrm{sys}}^4
[
\rho \mathcal{N}
+
(1-\rho)\mathcal{U}
]
}
\,d\Delta t.
\end{equation}
Eq.~\ref{final FIT} explicitly shows how accidental pair coincidences reduce the Fisher information available for clock-offset estimation. 
For any unbiased estimator satisfying regularity condition~\cite{steven1993fundamentals}, 
the lower bound on the SD is~$\sigma_{\text{CRLB},T}=\frac{1}{\sqrt{I_T}}$, such that~$\sigma_{\hat{\tau},T}\geq\sigma_{\text{CRLB},T}$,  where $\sigma_{\hat{\tau},T}$ is the SD of the experimentally estimated clock offset, $\hat \tau_T$, for the temporal-only estimator. Furthermore, $\sigma_{\text{CRLB},T}$ is tight for large $\rho$ but progressively loosens as $\rho$ decreases. As $\rho \rightarrow 0$, $\sigma_{\text{CRLB},T}$ tends to infinity, since $\tau$ is indistinguishable.

\vspace{0.5em}

\textit{(b) Polarization-assisted estimator}: After polarization post-selection, 
 the Fisher information is obtained from Eq.~\ref{final FIT} by replacing $\bar N_T$ and $\rho$ with $\bar N_P$ and $\rho_P$, respectively. Therefore, the Fisher information becomes
\begin{equation}\label{final FIP}
I_P
=
\bar N_P
\int_{0}^{\Delta \mathcal{T}}
\frac{
\rho_P^2
(\Delta t-\tau)^2
\mathcal{N}^2
}{
\sigma_{\mathrm{sys}}^4
[
\rho_P \mathcal{N}
+
(1-\rho_P)\mathcal{U}
]
}
\,d\Delta t,
\end{equation} and the corresponding lower bound on the clock-offset SD is~$\sigma_{\text{CRLB},P}=\frac{1}{\sqrt{I_P}}$.
We define the Fisher-information gain ratio as $G=\frac{I_P}{I_T}$. 

We next consider the strongly background-dominated limit of the photon-starved regime, for which $\bar N_a\gg \bar N_t$ and $q_a\bar N_a\gg q_t\bar N_t$. In addition, we assume that accidental pair coincidences dominate the timing distribution on $\Delta \mathcal{T}$, such that $\rho \mathcal{N}\ll(1-\rho)\mathcal{U}$ and $\rho_P \mathcal{N}\ll(1-\rho_P)\mathcal{U}$.
Under these conditions, $\rho \ll 1$ and $\rho_P \ll 1$, and therefore
$\rho \mathcal{N}+(1-\rho)\mathcal{U}\approx \mathcal{U}$,
and
$\rho_P \mathcal{N}+(1-\rho_P)\mathcal{U}\approx \mathcal{U}$.
Defining
$K=
\int_{0}^{\Delta \mathcal{T}}
\frac{
(\Delta t-\tau)^2
\mathcal{N}^2
}{
\sigma_{\mathrm{sys}}^4\mathcal{U}
}
\,d\Delta t$,
we obtain
$I_T\approx \bar N_T\rho^2K,
\
I_P\approx \bar N_P\rho_P^2K$ from Eq.~\ref{final FIT} and Eq.~\ref{final FIP}, respectively.
Note that the same $K$ appears in both estimators because polarization post-selection changes the retained number of pair coincidences and the true pair coincidence fraction, but not \(\mathcal{N}(\Delta t| \tau, \sigma_{\mathrm{sys}}^2 )\).
Since $\bar N_T \rho^2=\frac{\bar N_t^2}{\bar N_t+\bar N_a}$, under the condition $\bar N_a\gg \bar N_t$, we obtain~$I_T\approx\frac{\bar N_t^2}{\bar N_a} K$. 
Likewise,~$I_P\approx\frac{q_t^2\bar N_t^2}{q_a\bar N_a}K$. Therefore, in the photon-starved regime $G\approx\frac{q_t^2}{q_a}.$ Polarization post-selection therefore increases the Fisher information in the strongly background-dominated regime when $q_t^2>q_a$. 
At $\mathcal{V}=1$ and $q_a=\frac{1}{2}$, $G\approx2$, corresponding to $\frac{\sigma_{\text{CRLB},P}}{\sigma_{\text{CRLB},T}}\approx \frac{1}{\sqrt{2}}$. Thus, in the unit-visibility, unpolarized-background, strongly background-dominated limit, polarization post-selection approximately doubles the Fisher information and reduces the CRLB-limited clock-offset SD by approximately $30\%$. Conversely, when accidental pair coincidences are negligible and $\mathcal{V}=1$, polarization post-selection provides no Fisher-information advantage.
The analysis therefore predicts that the benefit of polarization post-selection emerges in the photon-starved regime. 
In the next section, we compare the experimentally estimated clock-offset SD with the CRLB based on true pair coincidences only, $\sigma_{\text{CRLB},t}$, defined from the measured true pair coincidence rate (defined in Sec.~\ref{experiment}).

\begin{figure}[ht] 
    \centering
    \includegraphics[width=1\linewidth]{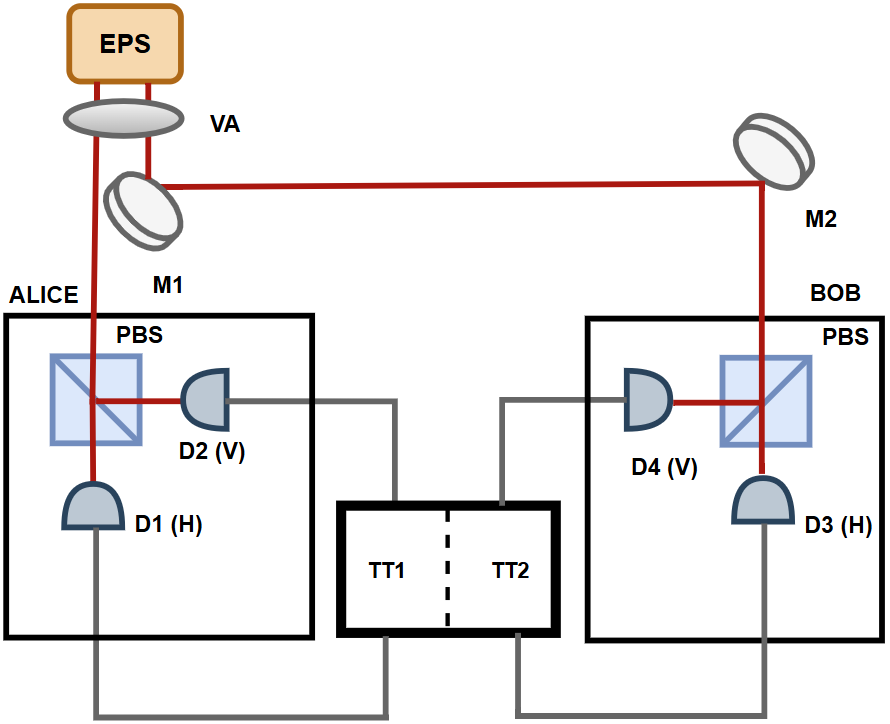}
    \vspace{4pt} 
    \caption{
    Experimental setup for the polarization-assisted clock-synchronization proof-of-concept.
    The entangled-photon source (EPS) generates polarization-entangled photon pairs, which are routed through the free-space optical channel using mirrors M1--M2.
    A variable attenuator (VA) introduces controlled channel loss symmetrically at Alice's and Bob's stations to emulate high-loss satellite-channel conditions.
    At Alice and Bob, polarizing beam splitters (PBSs) separate the horizontal (H) and vertical (V) polarization components, which are detected by single-photon counting modules D1--D4. See main text for the setup of TT1 and TT2.
    }
\label{fig. setup}
   \end{figure}

\section{Experimental proof-of-concept}\label{experiment}
Fig.~\ref{fig. setup} shows the experimental setup used to demonstrate the polarization-assisted clock-offset estimator. 
We used the polarization-entangled-photon source reported in~\cite{adhikari2026quantumsecuretimetransfer}. This source operates at a pump power of $1$mW, generates polarization-entangled photons centered at $811$nm, and has a measured H/V-basis polarization visibility, $\mathcal{V}$, of 0.99\footnote{$\mathcal{V}$ does not include polarization-reference-frame rotation error induced by satellite motion. This error can be largely compensated dynamically~\cite{yin2017satellite, wang2014experimental}. The residual polarization-reference-frame rotation error, i.e., the uncompensated component after dynamic compensation, would reduce $\mathcal{V}$ by only approximately $1\%$.}; the estimated SD of the intrinsic photon-pair arrival-time difference, $\sigma_{int}$, is~$1.5$ps. The entangled photons are distributed to Alice and Bob over separate $1.2$m free-space channels. 
At each receiver, a polarizing beam splitter separates the H- and V-polarized photons, which are detected by independent single-photon counting modules (SPCMs) with a dark-count rate of approximately~$100$s$^{-1}$ and a timing error,~$\sigma_{det}$, of~$250$ps. 
A Time Tagger Ultra (Swabian Instruments) with a timing error,~$\sigma_{tt}$, of 18ps records the SPCM outputs, including photon-arrival times and corresponding polarization outcomes, which together constitute the polarization-resolved detection data used for post-processing. We define an experimental `record' as the complete polarization-resolved detection data acquired over a single acquisition interval of duration $\Delta t_{\mathrm{acq}}=1\mathrm{s}$.

 Note that in the actual experiments we decided to emulate two time taggers (TT1 and TT2) using only one real time tagger - emulating the second tagger by adding an offset and drift to one set of timings. This allows us more freedom to explore the offset/drift parameter space. To this end, a controlled initial clock offset,~$\tau_0$, of $-5$ns, and a linear clock drift rate,~$\dot{\tau}$, of $10$ps/s are introduced in Bob's time tag arrays so that the clock offset evolves as $\tau_{\mathrm{}}(t')=\tau_0+\dot{\tau}t'$, where \(t' \in [0,\Delta t_{\mathrm{acq}}]\).
 The introduced $\dot{\tau}$ corresponds to a relative fractional-frequency offset of $|\Delta\nu/\nu_0|=|\dot{\tau}|=10^{-11}$, where $\nu_0$  is the nominal clock frequency and $\Delta\nu$ is the frequency difference between the two clocks. This value is consistent with that of two portable rubidium frequency standards~\cite{maurice2020miniaturized}. For the chosen $\Delta t_{\mathrm{acq}}$ value, $\sigma_{drift}\ll \sigma^{est}_{\mathrm{sys}}$ (terms discussed later).\footnote{The relative clock drift rate expected from two atomic clocks should have a marginal effect on the clock-offset estimation \textit{within a record}, therefore, explicit drift rate estimation is normally not absolutely necessary, and successive clock-offset estimates can be used to continuously update the clock, thereby tracking the accumulated effect of the clock drift rate over longer times.}  
The estimated system timing-error is~$\sigma^{est}_{\mathrm{sys}}= \sqrt{\sigma_{int}^2+2\sigma_{det}^2+2\sigma_{tt}^2 + \sigma_{drift}^2}\approx 354$ps, where $\sigma_{drift}$ is the clock-drift error, given by~$\frac{\dot{\tau}\Delta t_{\mathrm{acq}}}{\sqrt{12}} \approx3$ps.
 The corresponding true pair coincidence-only CRLB, $\sigma_{\text{CRLB},t}$, is determined to be $\sigma^{est}_{\mathrm{sys}}/\sqrt{R_t \Delta t_{\mathrm{acq}}}$, where $R_t$ is
the true pair coincidence count rate at Alice and Bob's photon detectors, given by $R_t=R_c-R_a$, where $R_c$ is the measured pair coincidence count rate, and $R_a$ is the accidental pair coincidence count rate estimated as~$S_1 S_2 \Delta t_c$, where \(\textit{S}_1\) and \(\textit{S}_2\) are the singles count rates measured at Alice and Bob, respectively, and $\Delta t_c$ is the coincidence window.

\textit{Emulated channel:}
We introduce controlled attenuation using a variable attenuator (VA) to emulate the high-loss conditions of a satellite quantum channel over the range of $40$--$70$dB, comparable to the combined loss experienced in the Micius experiments~\cite{lu2022micius} for transmitting and receiving telescopes with diameters of $0.3$m and $1.2$m, respectively, at an average separation distance of $1000$km. We define the total pair transmission efficiency as $\mathrm{\textit{T}}_{\text{ch}}=R_t/R_g$, where $R_g$ is the photon-pair generation rate from the source. 
The corresponding total pair loss is $\eta = 10\log_{10}\left(\frac{R_g}{R_t}\right)\,$dB, which we hereafter refer to as the total channel loss and which represents the combined loss of the Alice and Bob channels. Without loss of generality, we consider symmetric channels, such that each channel contributes $\eta/2$ to the total channel loss. Here, $\eta$ includes the fixed optical and detection losses, as well as the controlled attenuation introduced by the VA shown in Fig.~\ref{fig. setup}.
The value of $R_g$ is determined via $R_g=S_1S_2/R_t$~\cite{tanzilli2001highly} under low-background conditions, giving~$R_g\approx 4\times10^7$pairs~$\text{s}^{-1}$. 
We vary the laboratory illumination to obtain the desired background-count rate, $R_b$, measured by each SPCM. 
The measured accidental pair coincidences are approximately equally distributed among the HH, HV, VH, and VV outcomes, consistent with~$q_a \simeq \frac{1}{2}$.

We employ the two-stage coarse-to-fine synchronization algorithm from~\cite{rani2025obfuscatedquantumpostquantumcryptography} for clock-offset estimation.
For each record, we apply the algorithm twice to the same detection data. The temporal-only estimator uses only the photon-arrival times, whereas the polarization-assisted estimator additionally applies the proposed polarization post-selection before clock-offset estimation. 
The coarse alignment step searches over $\pm1$ms using $100$ns bins to localize the correlation peak. The fine alignment step then searches over $\pm100$ns around the coarse estimate using $200$ps bins, providing resolution below $\sigma^{est}_{\mathrm{sys}}$.

To evaluate synchronization performance, we acquire $N_{\mathrm{rec}}=100$ independent experimental records at each operating point. 
For fixed $\Delta t_{\mathrm{acq}}$ and synchronization parameters, operational synchronization-acquisition success depends on the distinguishability of the true pair correlation peak from the fluctuations of the accidental pair coincidence background. The true pair correlation peak amplitude is determined by $\sigma^{est}_{\mathrm{sys}}~\text{and}~R_t$, while increasing \(R_b\) increases the mean number of accidental pair coincidences in bin $\mathcal B_k$, \(\bar N_{a,k}\), and, under a Poisson approximation, its SD is \(\sqrt{\bar N_{a,k}}\). Polarization post-selection suppresses $\bar N_{a,k}$  to \(q_a \bar N_{a,k}\), lowering the SD to \(\sqrt{q_a \bar N_{a,k}}\) and thus improving peak distinguishability.
Following~\cite{PhysRevApplied.22.024012}, we quantify synchronization-acquisition performance by whether the estimated clock offset lies within a prescribed tolerance of the true clock offset.
For the estimator with  label \(l\in\{{T},{P}\}\),
we define the synchronization-acquisition indicator for the $m$-th experimental record as:
\begin{equation}
D_{l,m} =
\begin{cases}
1, & \left|\hat{\tau}_{l,m} - \tau_{\mathrm{true}}\right| \leq \Delta t_{w}, \\[4pt]
0, & \left|\hat{\tau}_{l,m} - \tau_{\mathrm{true}}\right| > \Delta t_{w},
\end{cases}
\label{eq:indicator}
\end{equation} where $\hat{\tau}_{l,m}$ is the experimentally estimated clock offset obtained using estimator $l$ for the $m$-th experimental record, $\tau_{\mathrm{true}}=\tau_0+\dot{\tau}\Delta t_{\mathrm{acq}}/2$ is the true clock offset at the midpoint of $\Delta t_{\mathrm{acq}}$, and $\Delta t_{w}$ is the synchronization-acquisition tolerance, taken to be $0.5$ns$~\approx1.5 \sigma^{est}_{\mathrm{sys}}$, providing a sufficiently narrow criterion for identifying successful acquisition of the true correlation peak. 
For $M$ experimental records, the synchronization-acquisition probability for estimator $l$ is~$\mathrm{P}_{\mathrm{succ},l} =
\frac{1}{N_{\mathrm{rec}}}\sum_{m=1}^{N_{\mathrm{rec}}} D_{l,m}$.
For each estimator \(l\), we also evaluate the experimental clock-offset SD,~$\sigma_{\hat{\tau},l}$, obtained over $M$ estimates \(\hat\tau_{l,m}\) at each operating point.

\begin{figure}[ht] 
    \centering
    \includegraphics[width=1\linewidth]{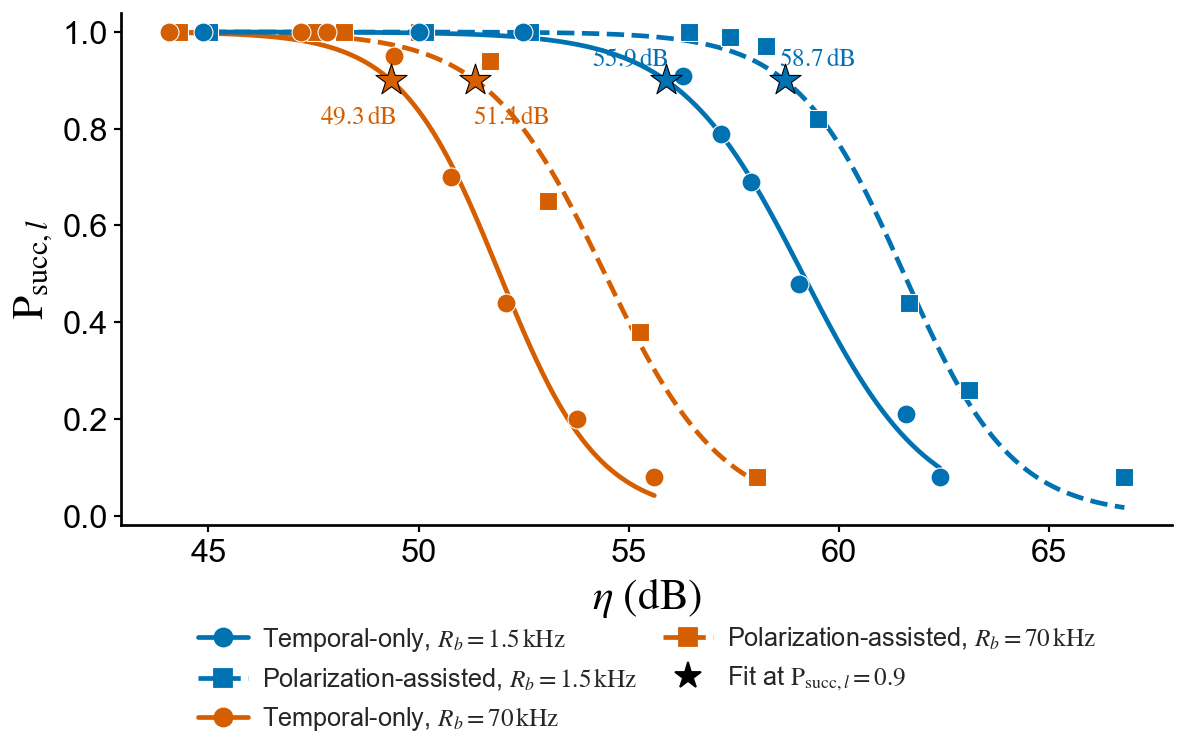}
    \vspace{4pt} 

    \caption{
Experimental synchronization-acquisition probability,~\(\mathrm{P}_{\mathrm{succ},l}\), as a function of total channel loss,~\(\eta\). 
Blue and orange denote background-count rates of \(R_b=1.5\mathrm{kHz}\) and \(R_b=70\mathrm{kHz}\), respectively. Curves are logistic least-squares fits. Stars denote the fitted synchronization-acquisition loss thresholds for~\(\mathrm{P}_{\mathrm{succ},l}=0.9\).}
\label{fig. experimental probability}
   \end{figure}

Fig.~\ref{fig. experimental probability} presents the experimentally extracted synchronization-acquisition probability for the temporal-only and polarization-assisted estimators at two background-count rates, $R_b=1.5$kHz and $R_b=70$kHz, representative of low-background and high-background operating conditions, respectively, as a function of $\eta$ ranging from $45$--$68$dB.
At lower total channel loss, both estimators reliably recover the clock offset. As the total channel loss increases, the temporal-only estimator transitions from reliable to unreliable acquisition, whereas polarization post-selection shifts this transition toward higher total channel loss. For the synchronization-acquisition criterion $\mathrm{P}_{\mathrm{succ},l}\ge0.9$, we define the synchronization-acquisition loss thresholds, $\eta_{90,T}(R_b)$ and $\eta_{90,P}(R_b)$, as the maximum tolerable total channel losses of the temporal-only and polarization-assisted estimators, respectively, at a given $R_b$.  At $R_b=70$kHz, the temporal-only estimator reaches $\eta_{90,T}=49.3$dB, whereas the polarization-assisted estimator reaches $\eta_{90,P}=51.4$dB, corresponding to a $2.1$dB extension in the synchronization-acquisition loss threshold. At $R_b=1.5$kHz, the corresponding thresholds are $55.9$dB and $58.7$dB, respectively, corresponding to an extension of $2.8$dB in the synchronization-acquisition loss threshold. Although \(q_a\simeq1/2\) at both background rates, the fluctuations in the accidental pair coincidence counts remain larger at higher \(R_b\), consistent with the smaller observed acquisition-loss extension. This nearly 2--3dB extension is operationally significant, as reliable synchronization acquisition can be maintained at up to approximately half the \(\mathrm{\textit{T}}_{\text{ch}}\) value required without polarization assistance. Satellite two-downlink channel loss can vary substantially throughout an orbital pass~\cite{lu2022micius}; consequently, a 2--3dB extension in the synchronization-acquisition loss threshold can potentially extend the portion of a pass over which synchronization remains reliably acquirable. For comparison, if this extension in the synchronization-acquisition loss threshold is attributed to an individual satellite-to-ground channel dominated by geometric spreading, it corresponds to approximately a $30\%$--$40\%$ increase in propagation distance.

\begin{figure}[ht] 
    \centering
    \includegraphics[width=1\linewidth]{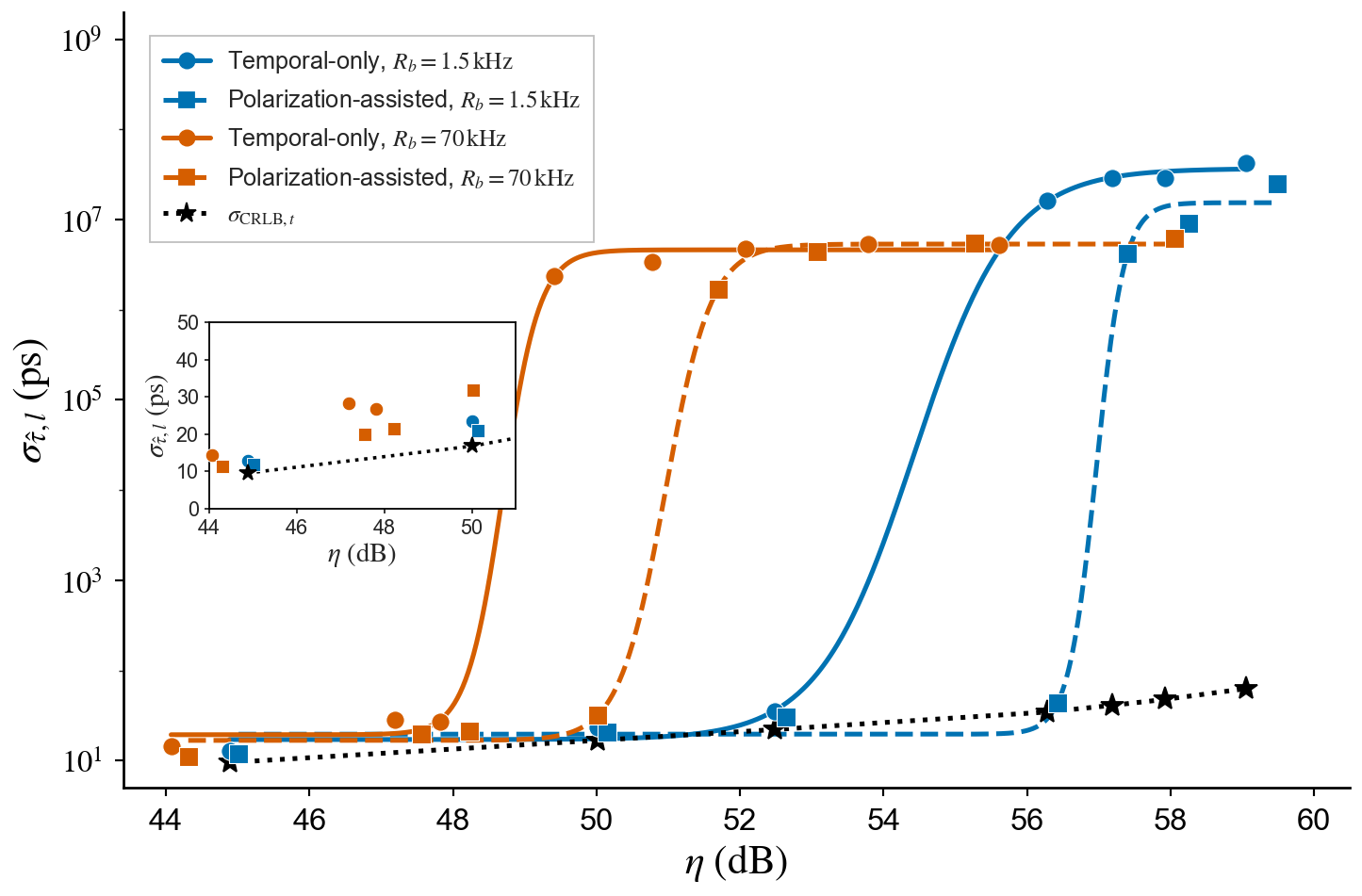}
    \vspace{4pt} 
\caption{
SD of the experimentally estimated clock offset, \(\sigma_{\hat{\tau},l}\), as a function of total channel loss, \(\eta\), for \(l\in\{T,P\}\). 
Blue and orange correspond to background-count rates of \(R_b=1.5\mathrm{kHz}\) and \(R_b=70\mathrm{kHz}\), respectively. The black dotted curve with star markers denotes the true pair coincidence-only CRLB, \(\sigma_{\text{CRLB},t}\). The inset zooms in on the low-loss region, where the experimentally estimated SDs are comparable to \(\sigma_{\text{CRLB},t}\). Curves are logistic fits to the experimental data. 
}
\label{fig: RMSE and spread}
   \end{figure}

Fig.~\ref{fig: RMSE and spread} shows the SD of the experimentally estimated clock offset, $\sigma_{\hat{\tau},l}$, along with the true pair coincidence-only CRLB. The polarization-assisted estimator achieves a lower SD than the temporal-only estimator, with the advantage becoming more apparent under high-background conditions. At lower total channel loss, the experimentally obtained SD approaches \(\sigma_{\text{CRLB},t}\) more closely.
At \((R_b=70\mathrm{kHz},\,\eta\simeq45\mathrm{dB})\), the temporal-only and polarization-assisted estimators yield \(\sigma_{\hat{\tau},T}=14.4\mathrm{ps}\) and \(\sigma_{\hat{\tau},P}=11.2\mathrm{ps}\), respectively, compared with \(\sigma_{\text{CRLB},t}\) of \(9.5\mathrm{ps}\). Therefore, polarization assistance reduces the SD by approximately \(22\%\) at \(45\mathrm{dB}\). Near \(\eta\simeq50\mathrm{dB}\), the polarization-assisted estimator maintains \(\sigma_{\hat{\tau},P}=31.8\mathrm{ps}\), whereas the temporal-only SD increases to \(2.35\times10^6\mathrm{ps}\), indicating synchronization-acquisition failure. This result shows that polarization assistance primarily extends the reliable synchronization-acquisition range to high $\eta$ values.

The remaining separation from \(\sigma_{\text{CRLB},t}\) shows that \(\sigma_{\mathrm{\hat{\tau}},l}\) remains above the theoretical lower bound, indicating scope for future improvements in estimator implementation. Nevertheless, the reduction in the SD with polarization assistance directly demonstrates its advantage in improving the clock-offset estimation.

\section{Simulation Study}

Having experimentally demonstrated $\mathrm{P}_{\mathrm{succ},l}$ over \(\eta=\)$45$--$68$dB at the two background count rate operating points, \(R_b=1.5\mathrm{kHz}\) and \(R_b=70\mathrm{kHz}\), we use Monte Carlo simulations to map $\mathrm{P}_{\mathrm{succ},l}$ over the wider operating region \(\eta=40\)--\(70\mathrm{dB}\) and \(R_b=0\)--\(300\mathrm{kHz}\), and to determine how polarization post-selection shifts the synchronization-acquisition boundary.

 \begin{figure*}[htp]
    \centering

    \begin{minipage}{0.48\linewidth}
        \centering
        \includegraphics[width=\linewidth]{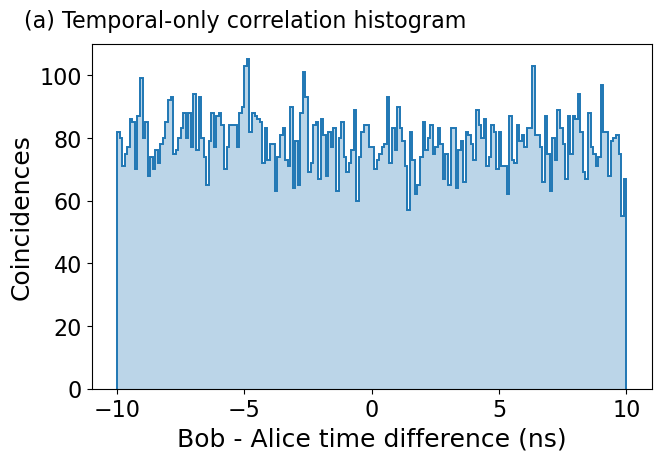}
    \end{minipage}
    \hfill
    \begin{minipage}{0.48\linewidth}
        \centering
        \includegraphics[width=\linewidth]{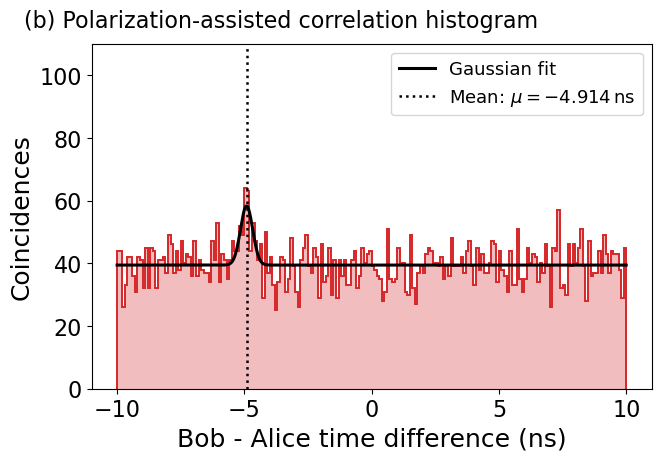}
    \end{minipage}

    \vspace{4pt}

    \caption{ Simulated temporal-correlation histograms illustrating synchronization acquisition failure and successful peak recovery under photon-starved conditions. (a) Temporal-only correlation histogram, in which the true pair correlation peak is obscured by accidental pair coincidences and cannot be reliably identified. (b) Polarization-assisted correlation histogram, where polarization post-selection suppresses mismatched accidental pair coincidences and reveals a distinct correlation peak near a delay of $-5$ns using a Gaussian-fit center, $\mu$, represented by the vertical dashed black line. The solid black curve shows a Gaussian fit to the recovered peak. }
    
    \label{fig:histogram}
\end{figure*}

\begin{figure*}[htp] 
    \centering
    \includegraphics[width=1\linewidth]{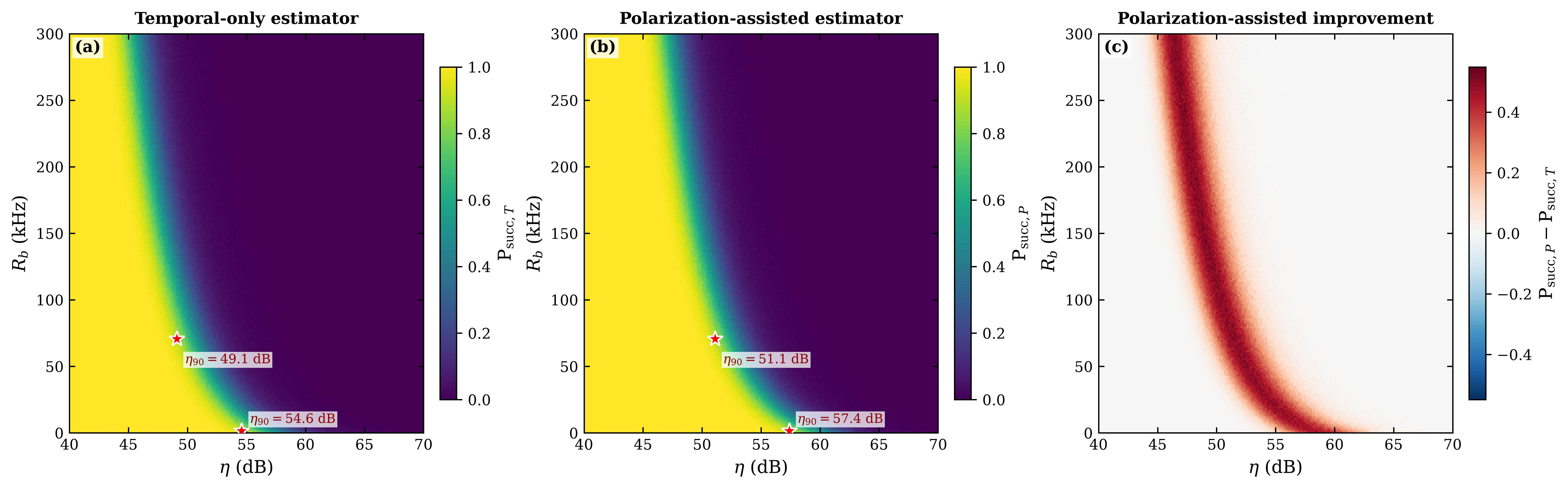}
    \vspace{4pt} 
    \caption{
Simulated synchronization-acquisition performance of the temporal-only and polarization-assisted estimators.
(a) Synchronization-acquisition probability, $\mathrm{P}_{\mathrm{succ},l}$, for the temporal-only estimator as a function of total channel loss, $\eta$, and background-count rate, $R_b$.
(b) Corresponding $\mathrm{P}_{\mathrm{succ},l}$ for the polarization-assisted estimator.
The marked points indicate the synchronization-acquisition loss thresholds for $\mathrm{P}_{\mathrm{succ},l}\geq 0.9$ at representative background-count rates.
(c) Difference in synchronization-acquisition probability, $P_{\mathrm{succ},P}-P_{\mathrm{succ},T}$, showing that the largest improvement occurs near the acquisition boundary.
}
\label{fig. acquisition_probability}
   \end{figure*}
\subsection{Simulation Details}\label{Simulation_detail}
In the simulation, we consider an entangled-photon source with a pair-generation rate of $4\times10^7$pairs~$\text{s}^{-1}$ and the same acquisition-interval duration, $\Delta t_{\mathrm{acq}}=1\mathrm{s}$, as in the experiment. For unit polarization visibility, $\mathcal{V}=1$, we construct  $2\times10^7$ time-tag pairs corresponding to the $\mathrm{HH}$ 
outcomes at Alice and Bob. We do likewise for the $\mathrm{VV}$ outcomes.
The time values associated with these time-tag pairs are uniformly distributed throughout the acquisition interval~$[0,\Delta t_{\mathrm{acq}}]$. The H and V resolved time-tag arrays at Alice are denoted by~$T_A^H$ and $T_A^V$, respectively, with~$T_B^H$ and $T_B^V$ defined likewise for Bob. 
An initial clock offset of $\tau_0=-5$ns and a linear clock drift rate of $\dot{\tau}=10\,\mathrm{ps/s}$, consistent with the experiment, are then introduced into Bob's time tags.
We then add independent Gaussian timing errors to Alice and Bob’s time tags. The timing errors are modeled as $\epsilon_A
\sim
\mathcal{N}(0,\sigma_A^2),~
\epsilon_B
\sim
\mathcal{N}(0,\sigma_B^2),$ with $\sigma_A=\sigma_B=250\mathrm{ps}$, where \(\sigma_A\) and \(\sigma_B\) are the corresponding SDs. Thus, the system timing-error SD is~$\sigma_{\mathrm{sys}}
=
\sqrt{\sigma_A^2+\sigma_B^2}
\simeq
354\mathrm{ps}$. 
Subsequently, we introduce $\eta$ by independently removing time tags from each time-tag array. In the simulation, \(\eta\) is varied from \(40\)--\(70\)dB. The loss is divided symmetrically between the two satellite-to-ground channels, such that each channel experiences $\eta/2$dB of loss. We add independent background events to each polarization-resolved time-tag array according to a Poisson process. The background-count rate in each time-tag array is denoted by $R_b$ and ranges from $0$--$300$kHz. 
Background events are generated independently and at equal rates in the $T_A^H$, $T_A^V$, $T_B^H$, and $T_B^V$ arrays. Consequently, the four accidental polarization combinations are equiprobable, and the probability that an accidental pair coincidence satisfies the polarization-matching criterion is $q_a=1/2$.

Finally, we concatenate \(T_A^H\) and \(T_A^V\) to form \(T_A\), while retaining the corresponding polarization labels in \(P_A\). We similarly construct \(T_B\) and \(P_B\) for Bob.
We use these arrays to estimate the clock offset. 
In the temporal-only estimator, we use only the $T_A$ and $T_B$ arrays; in the polarization-assisted estimator, we also use the $P_A$ and $P_B$ arrays.

\subsection{Simulation Results and Discussion}
We evaluate $\mathrm{P}_{\mathrm{succ},l}$ using the same criterion, $\Delta t_w=0.5\mathrm{ns}$, as in the experiment, for the temporal-only and polarization-assisted estimators as a function of $\eta$ and $R_b$ using $N_{\mathrm{MC}}=1000$ Monte Carlo trials at each operating point. In each trial, we use the same two-stage coarse-to-fine synchronization algorithm and parameter settings as used in the experiment to obtain the clock-offset estimates.

Fig.~\ref{fig:histogram} provides an illustrative example of the improvement provided by the polarization-assisted estimator. Under photon-starved conditions, accidental pair coincidences obscure the temporal-correlation peak, while polarization post-selection reduces the accidental pair coincidences and restores a clear peak centered at the true clock offset. The following results quantify this improvement over a wider range of channel conditions.
Fig.~\ref{fig. acquisition_probability} extends the experimentally observed behavior over a wide range of total channel loss and background-count rate. Figs.~\ref{fig. acquisition_probability}(a) and (b) show the synchronization-acquisition probability for the temporal-only and polarization-assisted estimators, respectively. Here, we again refer to the $\mathrm{P}_{\mathrm{succ},l}=0.9$ contour as the synchronization-acquisition boundary. At low total channel loss, both estimators reliably find the clock offset because the true correlation peak remains well above the accidental pair coincidence background. As the total channel loss and background-count rate increase, the temporal-only estimator transitions from reliable to unreliable synchronization acquisition. The polarization-assisted estimator shifts the acquisition boundary toward higher total channel loss for the same background-count rate.
Fig.~\ref{fig. acquisition_probability}(c)  shows that polarization post-selection primarily improves synchronization-acquisition probability near the acquisition boundary. This behavior is consistent with the analytical model (Fisher information): polarization post-selection provides little benefit when the temporal-correlation peak is already dominant, but becomes useful as the true-peak excess approaches the statistical fluctuations of the accidental pair coincidence background.
 At $R_b=70$kHz, the temporal-only estimator reaches $\eta_{90,T}=49.1$dB, whereas the polarization-assisted estimator reaches $\eta_{90,P}=51.1$dB, corresponding to a $2$dB extension in the synchronization-acquisition loss thresholds. At the lower-background operating point $R_b=1.5$kHz, the corresponding thresholds are $54.6$~dB and $57.4$dB, respectively, corresponding to an extension of approximately $3$dB in the synchronization-acquisition loss threshold. Therefore, the simulated results support the experimentally observed polarization-assisted estimator advantage over a broader operating region.

We note that the synchronization-acquisition improvement observed here originates from the polarization correlation rather than from quantum nonlocality itself; polarization entanglement is therefore not essential to the statistical post-selection advantage, provided sufficiently strong polarization correlations are available, although an entangled-photon implementation provides natural compatibility with quantum-communication and entanglement-verification protocols.

\section{Conclusion}
In this work, we proposed and experimentally demonstrated a polarization-assisted clock-offset estimator under photon-starved conditions relevant to satellite quantum channels. The proposed estimator uses the measured polarization correlations of entangled photon pairs to suppress accidental pair coincidences before clock-offset estimation. In the experimental proof-of-concept, the results demonstrate an extension  in the loss threshold of synchronization-acquisition, whose maximum value corresponds to an approximately $40\%$ increase in the propagation distance. 
Supporting Monte Carlo simulations show that the synchronization-acquisition advantage persists over a broader loss--background operating region. These results demonstrate that polarization information already available in detection data of entangled photons can improve synchronization robustness without requiring a separate synchronization signal.
Overall, the findings support the feasibility of polarization-assisted synchronization acquisition for high-loss, background-limited quantum channels relevant to satellite-to-ground communication.

\section*{Acknowledgments}
The authors thank Dr. Dushy Tissainayagam of Northrop Grumman Australia (NGA) for valuable discussions. This research was co-funded by NGA, the Defence Trailblazer (DT) Program, a collaborative partnership between the University of Adelaide and the University of New South Wales, and Australia’s Economic Accelerator (AEA)  Program. Both the DT Program and the AEA Program are supported by the Australian Government Department of Education.

{
\footnotesize
\bibliographystyle{IEEEtran}
\bibliography{IEEEabrv,main}
}

\end{document}